\documentclass[twocolumn]{aastex701}

\usepackage{amsmath,amssymb,amsfonts}
\usepackage{subcaption}
\usepackage{booktabs}
\usepackage{rotating}
\usepackage{overpic}
\usepackage{color}
\usepackage{graphicx}

\shorttitle{Planar structures in the $\Lambda$CDM universe}
\shortauthors{Jiang, Cen \& Peebles}
\submitjournal{ApJL}

\begin{document}

\title{Planar Structures in the $\Lambda$CDM Universe}

\author[orcid=0009-0001-2511-5862, sname=Jiang, gname=Xiaoqi]{Xiaoqi Jiang}
\affiliation{Institute of Astronomy, School of Physics, Zhejiang University, Hangzhou 310027, China}
\affiliation{Data Science Institute, The University of Chicago, Chicago, IL 60615, USA}
\email{xqjiang@uchicago.edu}

\author[orcid=0000-0001-8531-9536, sname=Cen, gname=Renyue]{Renyue Cen}
\affiliation{Institute of Astronomy, School of Physics, Zhejiang University, Hangzhou 310027, China}
\affiliation{Center for Cosmology and Computational Astrophysics, Institute for Advanced Study in Physics, Zhejiang University, Hangzhou, China}
\email[show]{renyuecen@zju.edu.cn}

\author[sname=Peebles]{P. J. E. Peebles} 
\affiliation{Joseph Henry Laboratories, Princeton University, Princeton, NJ 08544, USA}
\email{pjep@princeton.edu}

\begin{abstract}

The observed flat patterns in the distribution of galaxies around the Local Group raise a question:  
How common are such flat structures in the standard $\Lambda$CDM cosmological model?
We address this question by constructing mock observations from 
three independent large-box $N$-body simulations of the standard $\Lambda$CDM model. 
Using the same criteria as in Peebles (2023) 
we find a 0.3--0.4\% probability for a pair of Milky Way-sized galaxies 
in $\Lambda$CDM to be sitting on flat patterns 
as those observed at $z<0.021$. 
If allowance of 15 degrees of inclination is made between the $9.3$Mpc and
$93$Mpc planes, the probability increases to $\sim 1.5\%$.
Those lucky Local Group analogs stand out in that they are apparently markers of very rare density peaks of some sort, as their two-point correlation function has a correlation length of $\sim 60$Mpc, much higher than $12{\rm Mpc}$ for Milky Way--M31-like pairs not sitting on a 93~Mpc plane. 

\end{abstract}


\section{Introduction}

The primordial density fluctuations of the $\Lambda$CDM universe grow under gravity into the pattern of filaments, walls, and voids we observe today \citep{bondHowFilamentsGalaxies1996}.
Surveys have shown supercluster-scale structures emerging from this network, such as the Laniakea supercluster \citep{tullyLaniakeaSuperclusterGalaxies2014}, the CfA Great Wall \citep{GellerHuchra1989}, the Sloan Great Wall \citep{Gott2005}, and the BOSS Great Wall \citep{Lietzen2016}. 
Due to anisotropic collapse \citep{Zeldovich1970}, large-scale
sheets (or pancakes), where matter has collapsed along one axis, are common features and often associated with the superclusters.

The flattened distribution of local galaxies has been known for decades. 
\citet{deVaucouleurs1953} found that bright galaxies are concentrated toward a great circle on the sky, the planar system that he later termed the Local Supercluster \citep{deVaucouleurs1958}, and the plane itself was later formally defined as the supergalactic plane \citep{deVaucouleurs1976}. On smaller scales, the nearby galaxies lie in the thin Local Sheet that is close to and nearly parallel to this supergalactic plane, extending out to a distance of $7\rm\, Mpc$ \citep{Tully_2008,McCall2014}. 
Radio galaxies at $z\lesssim0.02$ \citep{ShaverPierre1989,Shaver1991} are found to be concentrated near the plane, and X-ray clusters define a slab aligned with it out to a radius of about $100\rm\, Mpc$ \citep{bohringerCosmicLargeScaleStructure2021}.
Beyond local regions, rich clusters across hundreds of megaparsecs \citep{Tully1986,Tully1987} are also found
to be aligned with this plane.
\citet{Peebles2022} reviewed this evidence and pointed out that in a region at least 170\,Mpc across
the clusters, radio galaxies, and most luminous early-type galaxies are close to this plane, while
spiral galaxies of similar luminosity are not.
\citet{peeblesFlatPatternsCosmic2023} further introduced a probe of thin flat patterns to scan counts of Milky Way-sized galaxies or larger in stacks of thin disks over the sky. Applied to galaxies at $z<0.021$, it has found a sheet, and the Local Sheet is a moderately bent part of it, inclined by $15^\circ$ to it. The continuation of this sheet is detected out to $z=0.085$ in disjoint redshift samples.

Simulation studies have so far considered the Local Sheet alone \citep{Goh2019,Neuzil2020,AragonCalvo2023}. However, how often an observer finds a local sheet that is embedded in, and nearly parallel to, a flat structure ten
times larger has not yet been measured, and it is unclear whether such patterns are consistent with the standard $\Lambda$CDM model.
We use three independent large-volume cosmological simulations of the standard $\Lambda$CDM model to test the presence of such arrangements. This article is presented as follows:
\S\ref{meth} describes the simulations, the probe and procedures used to analyze the simulations; \S\ref{results} presents the results, followed by concluding remarks in \S\ref{discussions}.

\section{Methodology}\label{meth}

\subsection{Simulations}

We use the $z=0$ snapshots of three independent simulations of box sizes
of $1$--$2h^{-1}$Gpc and dark matter particle masses of $5\times 10^8$--$2\times 10^9M_{\odot}$:
Uchuu \citep{ishiyamaUchuuSimulationsData2021,Oogi2023,dong-paezUchuuSDSSGalaxyLightcones2022,aungUchuuUniverseMachineDatasetGalaxies2022,pradaConfirmationStandardCosmological2023}, 
MDPL2 \citep{klypinMultiDarkSimulationsStory2016,Riebe2013,behrooziGravitationallyConsistentHalo2013}, and 
Flamingo-10K \citep{schayeFLAMINGOProjectCosmological2023,kugelFLAMINGOCalibratingLarge2023,Pizzati2024}. Table \ref{para} in Appendix \ref{simpars} 
summarizes the simulation details. 
These simulations have the desired properties of having large volumes and adequate mass resolutions for Milky Way-sized dark matter halos and larger.

Halo masses are $M_{200c}$ in Uchuu and $M_{vir}$ in MDPL2; for Flamingo-10K we convert $M_{200c}$ to 
$M_{102c}$ at $z=0$, which raises the masses by roughly one fifth and provides a comparison between definitions of halo 
mass (Appendix \ref{conver}). Flamingo-10K employs HBT-HERONS \citep{Han_2012,Han2018,moreno2025assessingsubhalofinderscosmological}, whereas 
the other two use Rockstar \citep{behrooziRockstarPhaseSpaceTemporal2013}. Small differences in the Hubble constant and box size among 
the datasets are standardized in the analysis.

\subsection{The luminosity--halo mass relation}\label{random_mapping}

We use dark matter halos as proxies for galaxies; halos that pass the thresholds set in Section~\ref{normalizations} are referred to as galaxies hereafter.
To relate the two, we adopt a relationship between galaxy luminosity and host halo mass by identifying $10^{12}M_{\odot}$ halos as Milky Way-like galaxies. In practice, we also introduce a scatter in this linear relation as 
\begin{equation}
    \log_{10} L = \log_{10}(\alpha\, M_h) + \epsilon\,,
    \label{eq:uncertainty}
\end{equation}
where $\alpha$ is a constant matching the mass-to-luminosity conversion and $\epsilon$
is a Gaussian random number with a standard deviation of 0.15~dex, consistent with \citet{Wechsler2018}. The normalization $\alpha$ is calibrated in 
Section~\ref{normalizations} by anchoring to the Milky Way. 

\begin{figure*}[!tp]
    \gridline{\fig{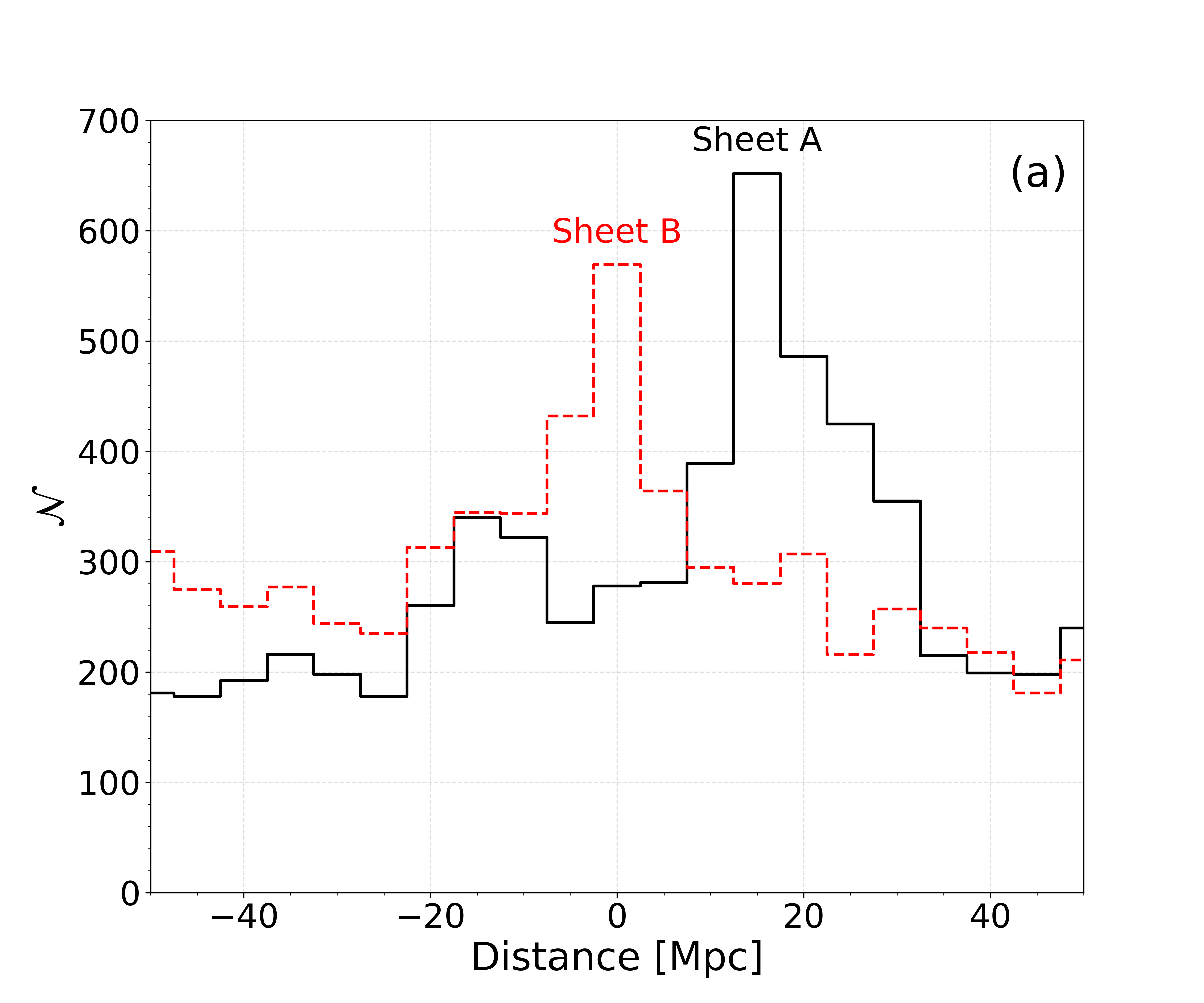}{0.42\textwidth}{}
              \fig{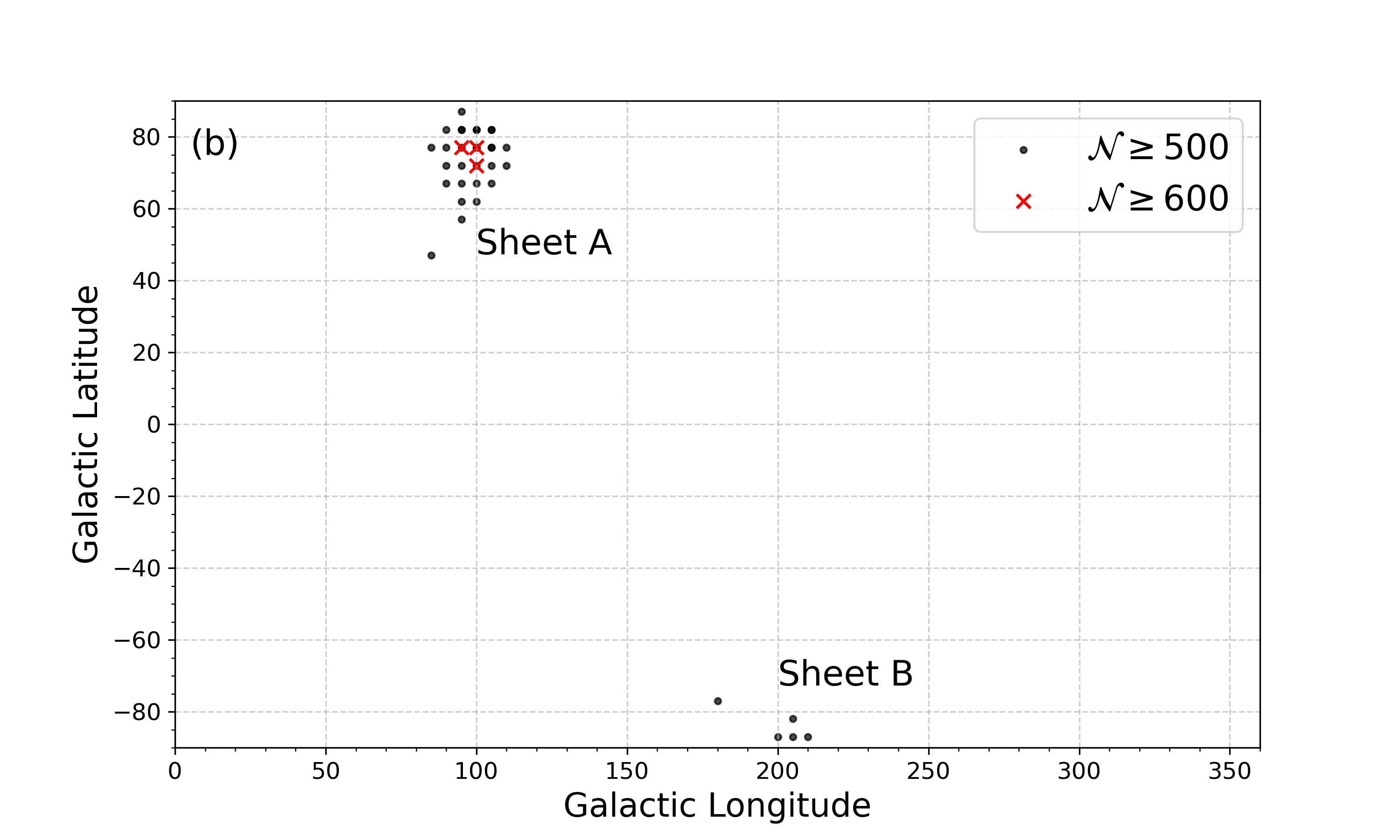}{0.58\textwidth}{}}
    \caption{Two directions in which a planar structure is detected in an LGA example. Panel (a): the normalized number distribution of galaxies along the direction at the center of a `sheet'. Panel (b): the angular distribution of `500-disks' (black dots) and `600-disks' (red crosses) over the sky. Each dot records the detection of a planar structure in that line-of-sight direction.}
    \label{example}
\end{figure*}

\subsection{Observers and probes}\label{hrp}

We select \emph{Milky Way-sized pairs} (hereafter Pairs) in the simulation, each consisting of two central galaxies with halo masses 
in the range of $1$--$2\times10^{12}\,M_{\odot}$,
separated by $0.7$--$1$~Mpc. 
We further filter them down to exclude those 
with a third galaxy whose halo mass is larger than $10^{12}\,M_{\odot}$ within 2~Mpc of 
either member. 
These Pairs form our parent sample; from their vantage points we place flat pattern probes employed by \citet{peeblesFlatPatternsCosmic2023} to quantify what fraction of these mock observers are able to find flat patterns we see in the Local Group, as described below.

The observer in a pair draws a unit direction vector $\vec{a}$ and places itself at the center of a cylindrical disk of a thickness $h$ and radius $r$ equal to 
\begin{equation}
    \begin{aligned}
    h&= 5\,\text{Mpc} \\
    r&= 75\,\text{Mpc}\,,
    \end{aligned}
\end{equation}
representing a flat probe, whose normal is $\vec a$.
Stacked along the same vector direction it further places 10 more disks  of the same size contiguously, reaching $z=0.021$ at the farthest disk.
The same is done in direction $-\vec a$, resulting in a stack of 21 disks altogether.
The number of relevant galaxies within each disk is recorded, which then forms a \emph{higher-redshift probe},
as in \citet{peeblesFlatPatternsCosmic2023}.

Once we are done with one vector $\vec a$,
 we scan the entire 4$\pi$ sky on the spherical Fibonacci lattice \citep{Hannay2004, Gonzalez2010}, a deterministic, quasi-uniform point set in which each direction subtends nearly the same solid angle. It avoids both the polar crowding of equiangular grids and the slow $\sqrt{N}$ convergence of random sampling. 
 Our angular resolutions range from $1296$ directions per hemisphere at the coarsest to approximately $10^6$ at the finest; with 21 disks per direction, the finest sampling yields $\approx 2\times10^7$ disk counts per observer per probe scale. Convergence with respect to angular resolution is demonstrated in Appendix \ref{asym}.

A second, scaled-down set of \emph{nearby probes} measures the small-scale environment of the same observers. These use the identical 21-disk geometry with
\begin{equation}
    \begin{aligned}
        h &= 0.5\,\text{Mpc} \\
        r &= 7.5\,\text{Mpc}\,,
    \end{aligned}
\end{equation}
reaching 9~Mpc along the normal of the disks from the observer, equivalent to $z \approx 0.0021$. The two scales cover a decade in distance and serve complementary 
purposes: \emph{higher-redshift probes} identify extended planar structures, and nearby probes characterize the local matter distribution.

\subsection{Normalization}\label{normalizations}

Normalization of galaxy counts to fit the magnitude limit of the observed catalogs proceeds in two steps: a global step that fixes the halo mass thresholds, and a local step that defines high-density disks.

The global step is a calibration against the observed count. 
For \emph{higher-redshift probes} 
we select a large number of random 
spheres and adjust the lower halo mass threshold until the mean number of galaxies within the 
$z=0.021$ spheres reaches 8540, matching the number in the observational sample 
\citep{peeblesFlatPatternsCosmic2023}. The resulting cutoff is very close to $10^{12}\,M_{\odot}$, consistent with the 
survey magnitude limit $M_{K_s}\approx-23$; therefore, the simulated samples turn out to be normalized to the observations at 
the scale of Milky Way-mass halos.
For \emph{nearby probes}, calibration by counts is no longer reliable on scales of $\sim 10$~Mpc, as the local galaxy distribution is modulated by the Local Void \citep{Tully1988, Peebles2010}. We instead anchor the luminosity--halo mass relation at the Milky Way, $M_{B,\rm MW} = -20.9$ with the fiducial Milky Way halo mass as $M_{h,\rm MW} = 10^{12}\,M_\odot$.
The absolute magnitude assigned to a halo of mass $M_h$ is then
\begin{equation}
    M_B = -20.9
        - 2.5\,
          \log_{10}\!\left(\frac{M_h}{10^{12}\,M_\odot}\right)
        - 2.5\,\epsilon\,,
    \label{eq:mag_halo}
\end{equation}
with $\epsilon$ as in Equation~\ref{eq:uncertainty}. We neglect the additional scatter from the variation of the $B$-band mass-to-light ratio across galaxy types \citep{Bell2003},
and exclude halos below $10^{10}\,M_{\odot}$ as insufficiently resolved.

For \emph{higher-redshift probes} we first define two statistics,
(threshold 1) $\text{T}_1$ and (threshold 2) $\text{T}_2$, by counting galaxies in each of the large disks defined as 
\begin{equation}
    \begin{aligned}
    \text{T}_1 = 500 \times \frac{N_{sphere}}{8540} \\
    \text{T}_2 = 600 \times \frac{N_{sphere}}{8540}\,,
    \end{aligned}
    \label{threshold_definition}
\end{equation}
where $N_{sphere}$ is the number of galaxies in the sphere
of radius $93$Mpc.
If the number of galaxies within any of the disks that an observer in a pair finds after scanning the whole sky exceeds $\text{T}_1$ ($\text{T}_2$), 
this observer is said to have a 500-disk (600-disk),
as defined by \citet{peeblesFlatPatternsCosmic2023}. 
Here the accounting is performed in redshift space, conforming to observations.

For \emph{nearby probes},
due to the small number of galaxies within 10~Mpc, we require that the most populated disk with $N_{max}$ galaxies
contain a significant fraction of the total within the sphere,
\begin{equation}
    \frac{N_{max}}{N_{0.0021}} > \frac{12}{53}\,,
    \label{small_1}
\end{equation}
where $N_{0.0021}$ is the number of galaxies within the local sphere of radius 9~Mpc. Around the Milky Way, 13 of the 53 galaxies lie in the central disk \citep{peeblesFlatPatternsCosmic2023}, and requiring the fraction to exceed 12/53 lets the Local Group itself satisfy the condition. Disks satisfying such a condition are called `small-disks'.

As a check on the simulation data, Appendix \ref{simpars} verifies that the mass distributions are isotropic to the precision we need.
The observed planar structures around the Local Group have a particular angular arrangement, which we adopt as the pattern to search for. 

\begin{figure*}[!tp]
    \gridline{\fig{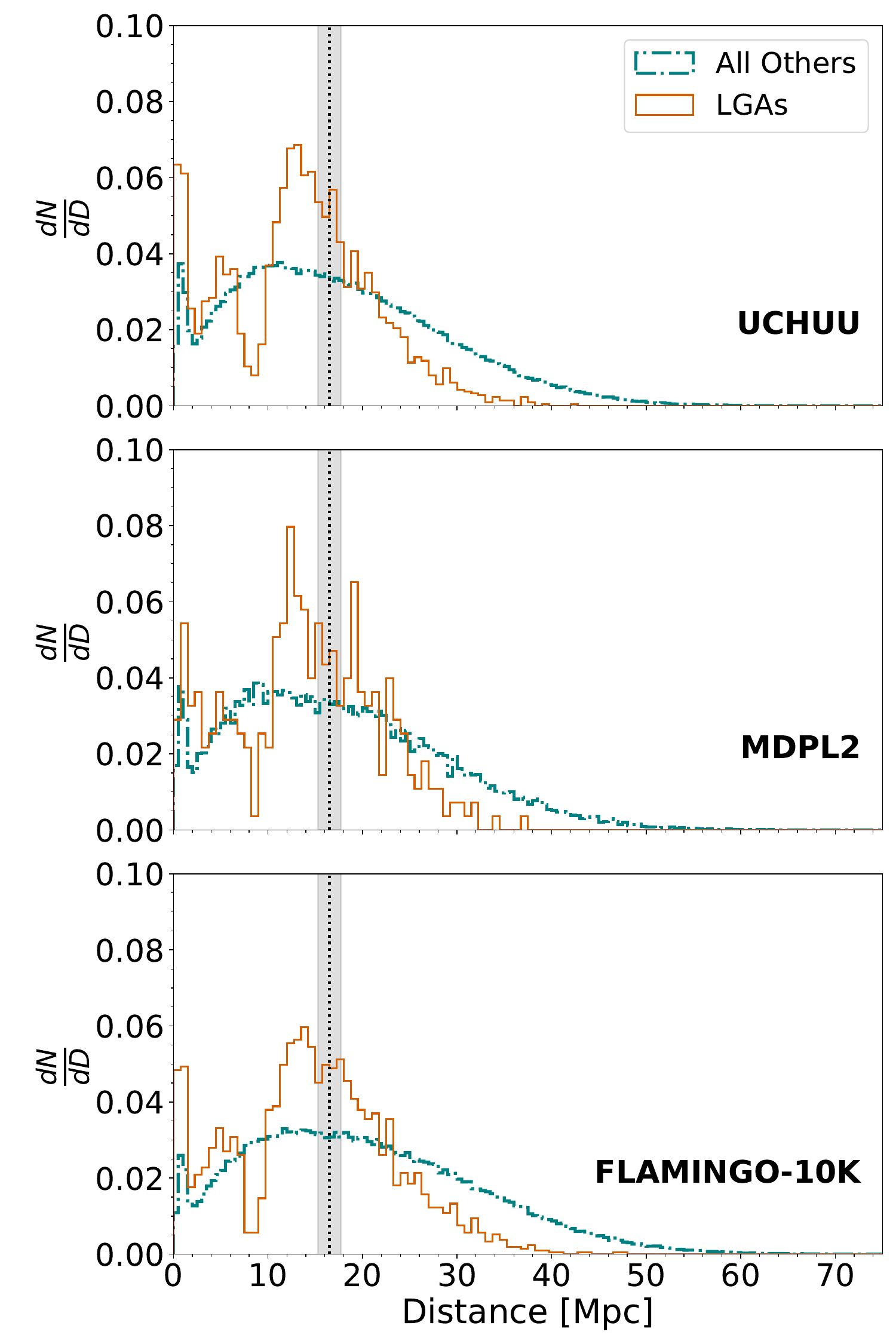}{0.275\textwidth}{(a)}
              \fig{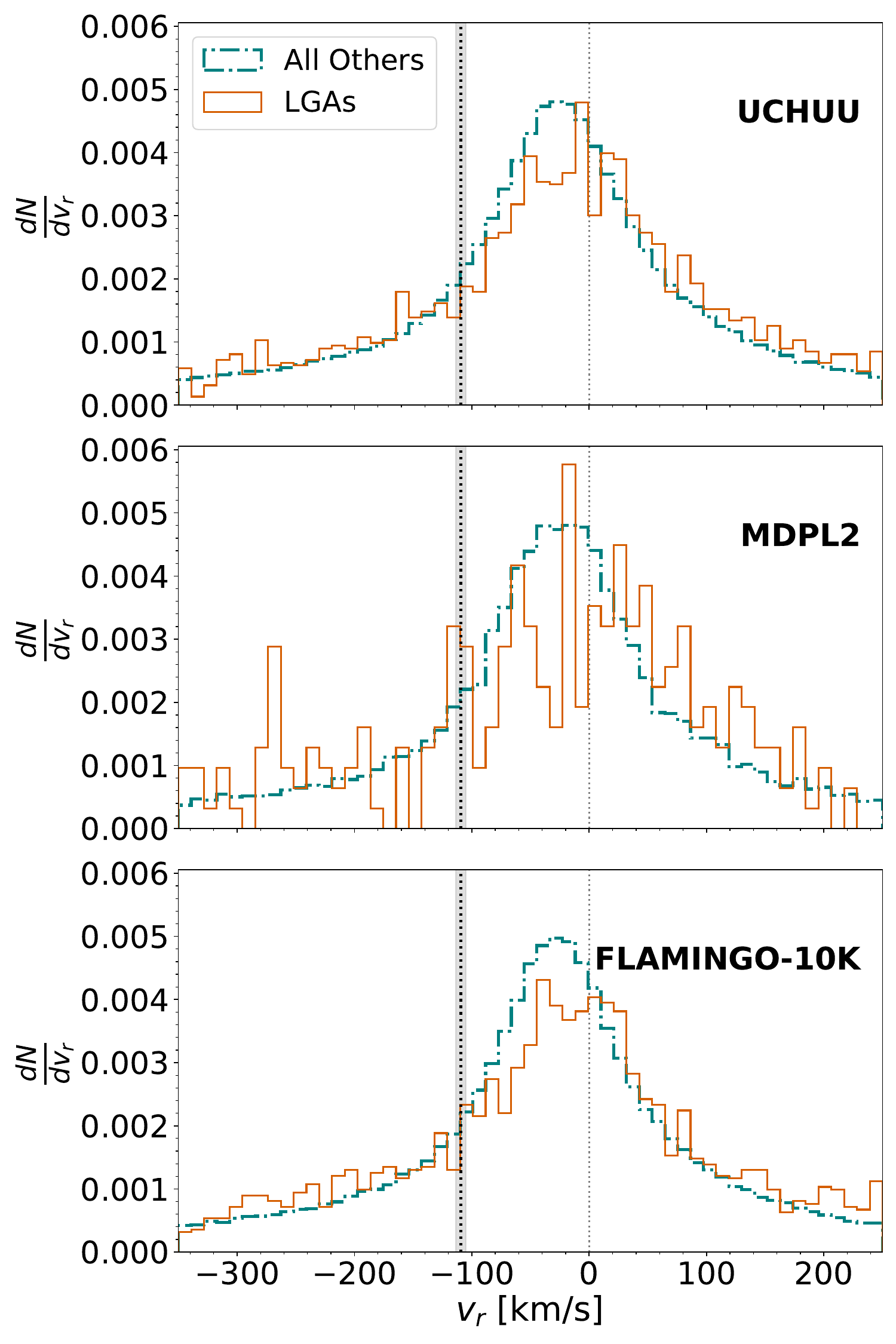}{0.275\textwidth}{(b)}
              \fig{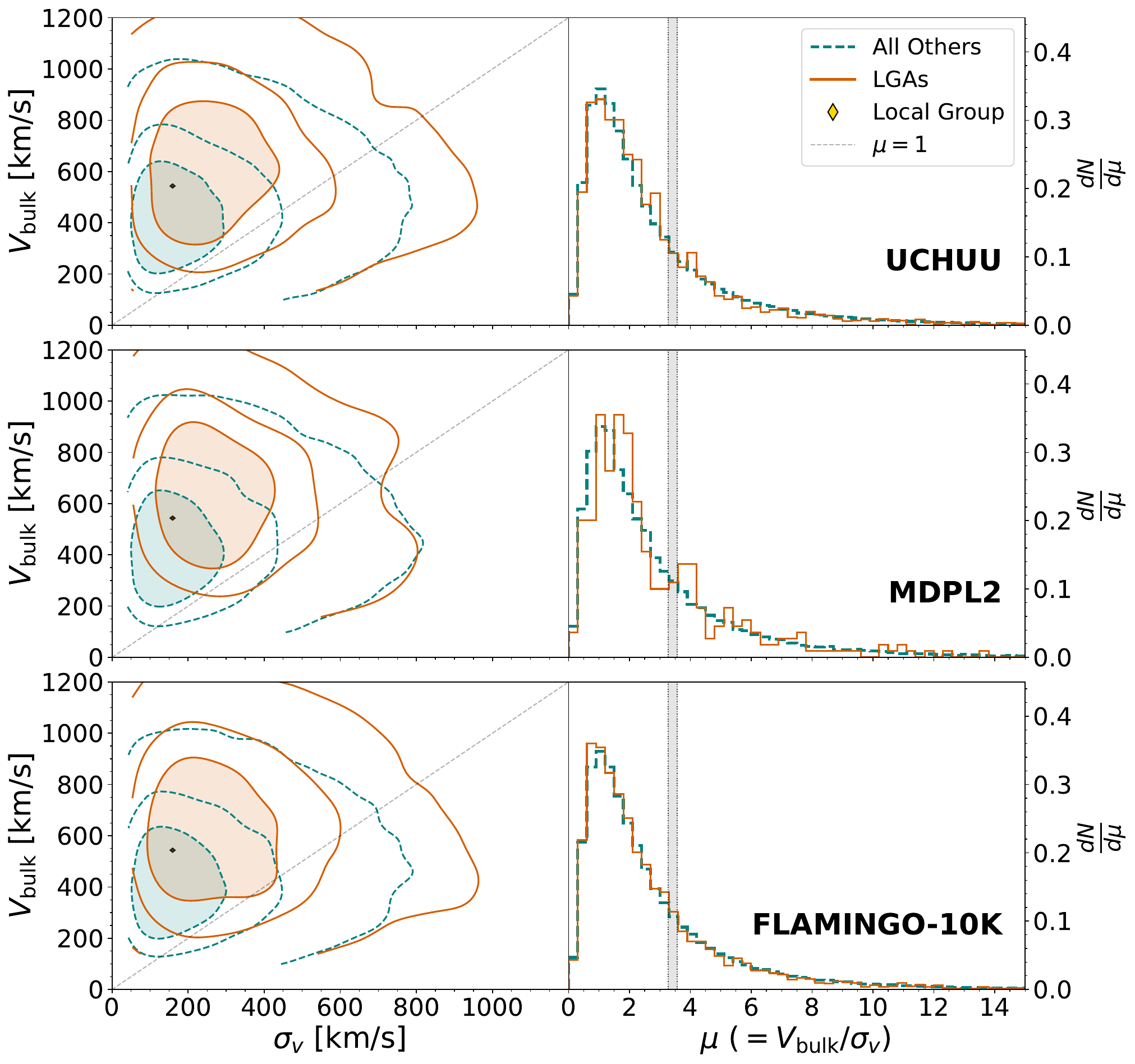}{0.44\textwidth}{(c)\hspace{0.18\textwidth}(d)}}
    \caption{Environments of LGAs compared with those of ordinary (all other) Pairs. 
    Panel (a): distances from Pairs to the nearest cluster more massive than $10^{14}M_{\odot}$; orange marks LGAs, cyan all other Pairs, and the dotted line with gray shading the Local Group--Virgo distance with its uncertainty as a rough reference \citep{Mei_2007}. 
    Panel (b): relative radial velocities between two members of each pair (Hubble flow is added for simulated galaxies); the gray band shows the relative velocity of the MW--M31 pair from the measured velocity \citep{vanderMarel2012}.
    Panel (c): bulk flow velocity against velocity dispersion within a local 7~Mpc volume; contours enclose 38\%, 68\%, and 95\% of the probability density, and diamonds mark the values for our own Local Group under luminosity cuts $M_K < -22.5$ and $M_K < -23.5$ \citep{Karachentsev2013}. 
    Panel (d): the corresponding cosmic Mach numbers; the gray band shows the range estimated for our neighborhood under the same cuts.}
    \label{fig:env}
\end{figure*}

\subsection{Selection of LGAs}\label{LGA}

From the Pairs selected, 
we define those that sit on flat structures 
similar to those around the real Local Group as Local Group analogs (LGAs)
when they meet the following conditions.
\begin{enumerate}
    \item \textbf{Valid Sample:} The pair's $z < 0.0021$ sphere contains at least 40 galaxies. This bounds the denominator of Equation \ref{small_1} to avoid small number-induced noise; this criterion removes about one fifth of the Pairs, and all subsequent additional selections are based on this valid sample.
    \item \textbf{Large Disk:} A 600-disk is required, surrounded by multiple 500-disks within a cone of half-angle $15^\circ$.
    \item \textbf{Local Disk:} At least one small-disk is required as the counterpart of the Local Sheet.  
    \item \textbf{Alignment between Local and Large Disks:} The LGA sits on a small flat structure which is embedded within and parallel to a large flat structure. 
    We allow the large and small disks to be inclined to as much as 15 degrees in the most lenient case.
\end{enumerate}

The Pairs in the valid sample that are not LGAs are hereafter referred to as general (or all other) Pairs.

Figure \ref{example} shows the galaxy counts for a random LGA. Panel (a) resembles the counts observed in the local universe \citep{peeblesFlatPatternsCosmic2023}: because the system lies within or near a plane, the counts are enhanced in the central bins. The large-scale geometries of these systems are also comparable to that of the Local Group.

\section{Results}\label{results}

\subsection{Occurrence Rate of LGAs in the $\Lambda$CDM model}

We require the observer to lie in the central disk on both scales, and consider two directional couplings between the two planes: exactly parallel ($0^\circ$) or inclined by at most $15^\circ$, thereby constructing two configurations, $1\mathrm{mid}+1\mathrm{mid}+0^\circ$ and $1\mathrm{mid}+1\mathrm{mid}+15^\circ$, where 1mid denotes that the observer lies in the central disk of the probe. Table~\ref{prob} summarizes the results.

\begin{deluxetable*}{llrrrrrrr}
\tabletypesize{\scriptsize}
\tablewidth{0pt}
\tablecaption{Measurements for LGAs \label{prob}}
\tablehead{
\colhead{Configuration} &
\colhead{Simulation} &
\colhead{Sample} &
\colhead{Valid} &
\colhead{Local} &
\colhead{600-Disk} &
\colhead{Large} &
\colhead{LGA} &
\colhead{Fraction} \\
\colhead{} &
\colhead{} &
\colhead{Size} &
\colhead{Sample} &
\colhead{Disk} &
\colhead{Existence} &
\colhead{Disk} &
\colhead{} &
\colhead{(\%)} \\
\hline
\colhead{(1)} & \colhead{(2)} & \colhead{(3)} & \colhead{(4)} &
\colhead{(5)} & \colhead{(6)} & \colhead{(7)} & \colhead{(8)} & \colhead{(9)}
}
\startdata
$1\mathrm{mid}+1\mathrm{mid}+0^\circ$  & Uchuu         & 317910 & 246861 & 130739 & 29709 & 25960 & 697  & 0.282(0.331) \\
                                   & MDPL2         & 38136  & 30056  & 15461  & 3920  & 3428  & 65   & 0.216(0.294) \\
                                   & Flamingo-10K  & 246726 & 191903 & 103861 & 26362 & 22911 & 673  & 0.351(0.412) \\
\hline
$1\mathrm{mid}+1\mathrm{mid}+15^\circ$ & Uchuu         & 317910 & 246861 & 130739 & 29709 & 25960 & 2817 & 1.141(1.318) \\
                                   & MDPL2         & 38136  & 30056  & 15461  & 3920  & 3428  & 368  & 1.224(1.490) \\
                                   & Flamingo-10K  & 246726 & 191903 & 103861 & 26362 & 22911 & 2811 & 1.465(1.661) \\
\enddata
\tablecomments{Candidate counts and statistics for the two configurations. Column (1) gives the angular coupling between the local and large planes; 1mid denotes that the observer lies in the central disk at both scales. ``Sample Size'' is the number of Pairs in the parent sample; ``Valid Sample'' lists samples 
satisfying the local number density threshold ($\ge40$); ``Local Disk'' lists those passing the 
small-scale planarity test; ``600-Disk Existence'' lists those containing at least one 600-disk; 
``Large Disk'' lists those hosting at least one Large Flat Structure meeting the 500-disk criterion 
around a 600-disk. Fraction = $\#(\rm LGA)/\#({\rm Valid\ Sample})$. Values in parentheses are 
asymptotic results as the angular sampling resolution approaches infinity (Appendix \ref{asym}).}
\end{deluxetable*}

For exactly parallel planes ($1\mathrm{mid}+1\mathrm{mid}+0^\circ$), the probability is 
0.29--0.41\%.
The configuration with the inner disk inclined
by at most 15 degrees to the outer disk 
($1\mathrm{mid}+1\mathrm{mid}+15^\circ$) yields a probability of 
1.32--1.66\%.
Readers should bear in mind that the Local Group possesses two independent 600-disks along distinct directions and at different distances, so these probabilities are conservative upper bounds. 

\begin{figure*}[!tp]
    \centering
    \includegraphics[width=0.8\textwidth]{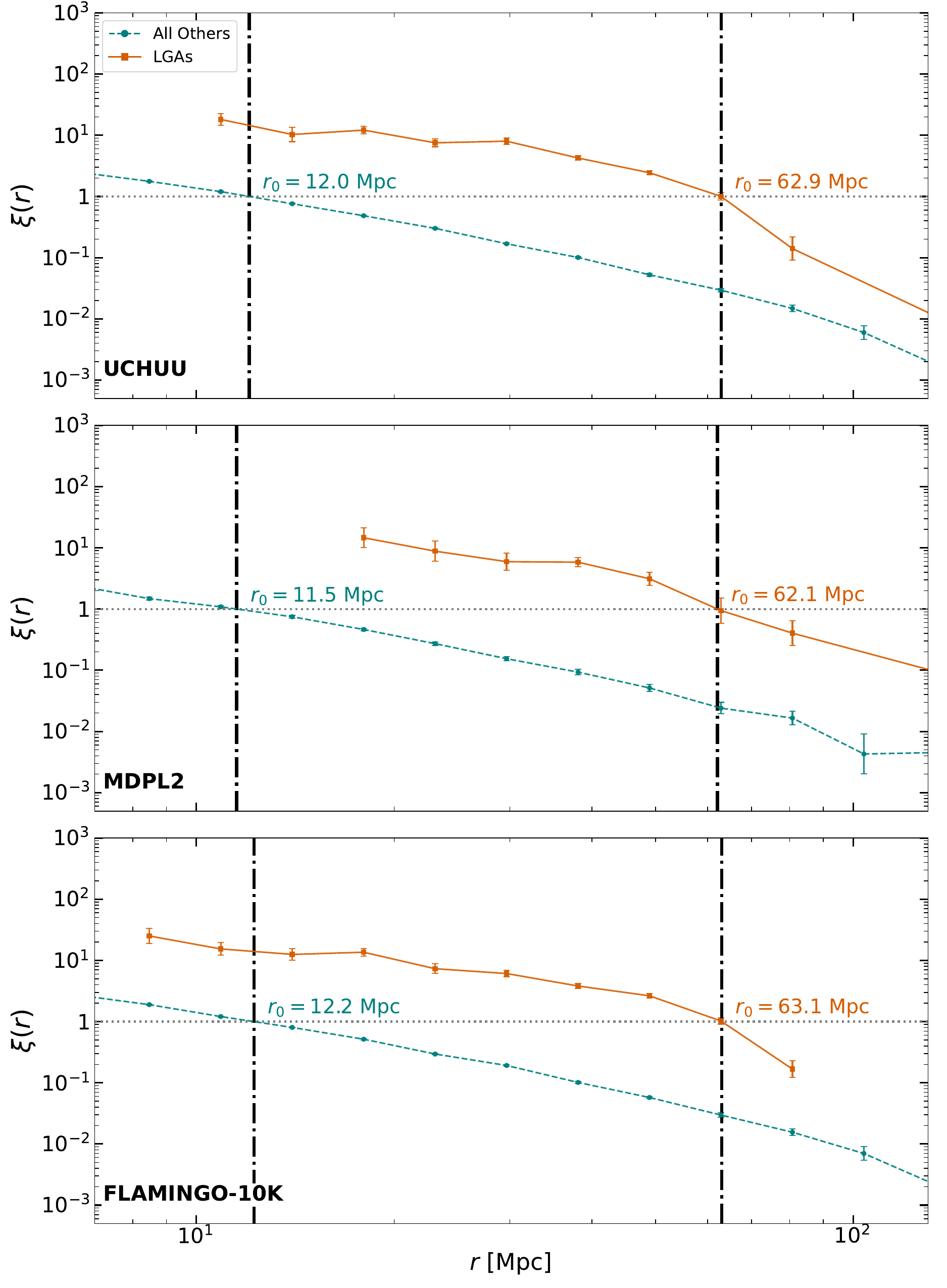}
\caption{Two-point correlation functions of LGAs (orange solid) and of the general Pairs (cyan dashed). Vertical dash-dotted lines mark the crossings $\xi(r_0)=1$ obtained by $\log$-$\log$ linear interpolation between the last bin above $\xi=1$ and the next bin below. Error bars are from 27-fold jackknife resampling. We only show bins with expected random pair count $\langle RR\rangle\geq0.3$, which removes the smallest separations for LGAs, where too few pairs are expected; bins with $\xi\leq0$ (LGAs beyond about 100 Mpc) cannot be shown on the logarithmic axis.}
    \label{corr}
\end{figure*}

\subsection{Are LGAs special in some ways?}

We test whether LGAs differ statistically from the general Pairs not sitting in such arrangements. Since the $1\mathrm{mid}+1\mathrm{mid}+0^\circ$ configuration holds too few LGAs to generate reliable statistics, we only present the results from the $1\mathrm{mid}+1\mathrm{mid}+15^\circ$ configuration.
Figure \ref{fig:env}(a) compares the distances to the nearest massive cluster (halo mass above $10^{14}M_{\odot}$), and shows that LGAs lie systematically nearer to clusters than the general Pairs.

Figure~\ref{fig:env}(b) shows the distribution of the relative radial velocity between the two members of each pair in the valid sample. This velocity is the difference of their peculiar velocities projected onto the separation direction, plus the Hubble flow $H_0$ times their separation, so it can be compared directly with the observed MW--M31 value. Differences between LGAs and the general Pairs are small, showing that proximity to such planar structures
does not, as a whole, affect the dynamics between the largest members of a small local configuration.
The relative velocity between MW and M31 of about $-110$~km\,s$^{-1}$ is consistent with the distribution, but not at the peak.

We further quantify the local velocity field with the Cosmic Mach Number \citep{1990ApJ...348..378O},
the ratio of the bulk flow to the velocity dispersion within a region.
This ratio is computed for galaxies above the same mass threshold used for the higher-redshift probes. Figure \ref{fig:env}(c) shows the bulk flow against the dispersion within a 7~Mpc (approximately $5\,h^{-1}$Mpc) box, and Figure \ref{fig:env}(d) the resulting Mach numbers. For comparison we compute the same quantities for our own neighborhood from the catalog of \citet{Karachentsev2013}, with absolute magnitude cuts $M_K < -22.5$ and $M_K < -23.5$ chosen to encompass galaxies of Milky Way size or larger. The Mach number distributions of LGAs and the general Pairs are nearly identical, due to the correlation between the bulk flow velocity and dispersion for the LGAs (Figure~\ref{fig:env}(c)). Our own Local Group is fairly typical in this measure.  Nonetheless, it is noted that the bulk of LGAs or the general Pairs are in regions of Mach number of about unity.

\subsection{What density fluctuations do LGAs represent?}

\begin{figure}[!htbp]
    \centering
    \includegraphics[width=\linewidth]{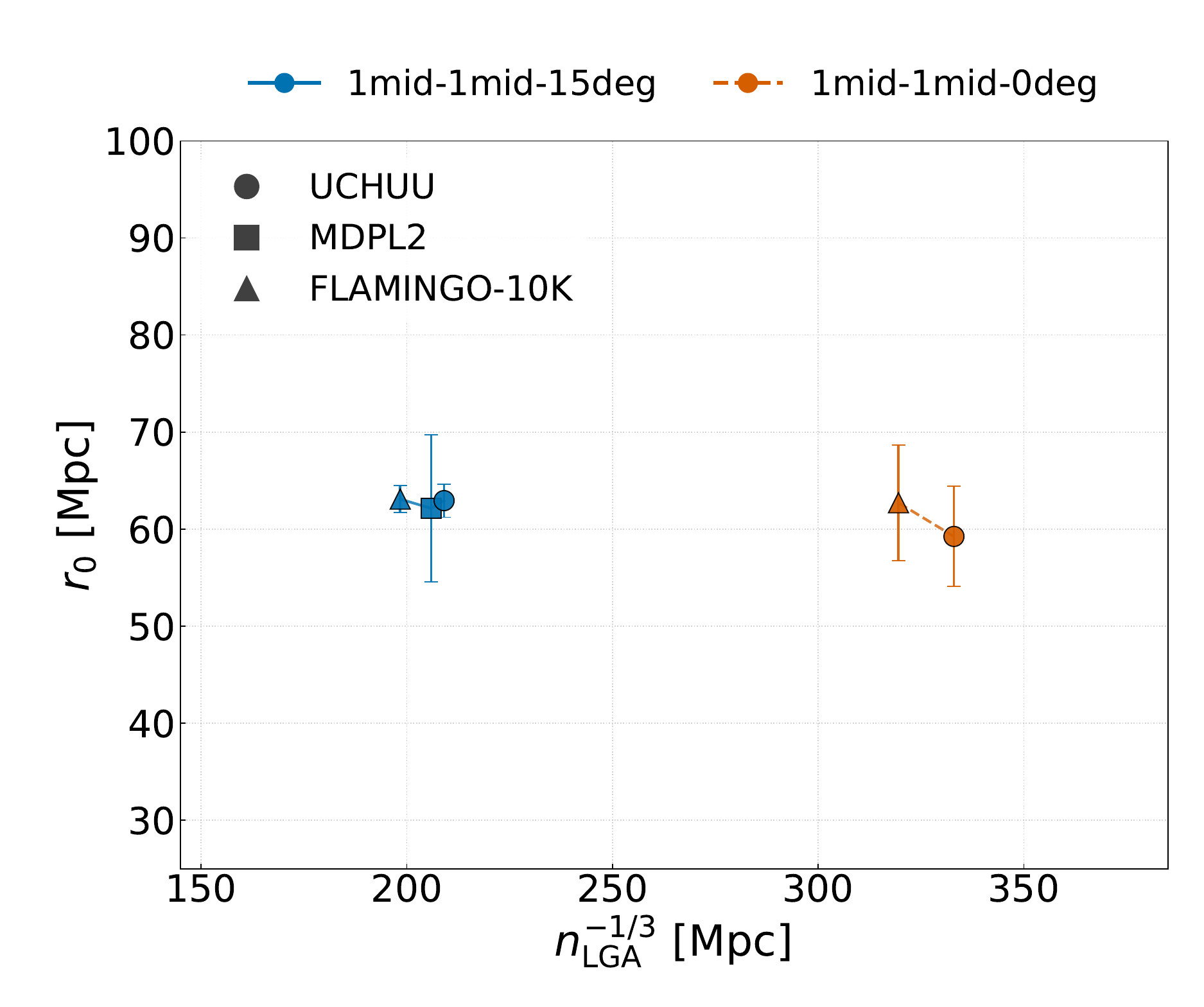}
    \caption{
        Correlation length across the two configurations and simulations as a 
        function of number density of corresponding LGAs. 
        Error bars are jackknife errors computed with the same method as in Figure \ref{corr}.
The MDPL2 point for $1\mathrm{mid}+1\mathrm{mid}+0^\circ$ is not shown because its sample is too sparse for $\xi(r)$ to cross unity.
    }
    \label{consist}
\end{figure}

We estimate the correlation function in each separation bin as
$\xi=DD/\langle RR\rangle-1$, where $DD$ is the number of pairs in the sample with
separations in bin $k$, $[r_k,r_{k+1})$, and
$\langle RR\rangle=\frac12N(N-1)\,\frac{4\pi}{3}(r_{k+1}^3-r_k^3)/L^3$ is the number
expected for the same $N$ points placed at random in the periodic box of side $L$.
Figure \ref{corr} shows the two-point correlation functions of LGAs and of the general Pairs that are computed as above.
The curve for the general Pairs is smoother owing to its larger sample. The LGA two-point correlation function gives a correlation length of $\sim 60$Mpc, 
consistent across three simulations, and stable across two configurations, as shown in Figure \ref{consist}.
This is a very large correlation length,
compared with, for example, of order $20-30$Mpc for rich clusters.
This indicates that LGAs represent some rare events in the primordial Gaussian random field, but not necessarily density peaks in the conventional sense that form collapsed, virialized structures later.
We speculate that these may be very rare events in lower-dimensional space, as opposed to spherical/ellipsoidal density peaks.
Of course, LGAs are merely markers of these structures, not structures themselves. We defer a physical understanding of the origin to a future study.

\section{Discussion and Conclusions}\label{discussions}

We use $N$-body simulations to examine whether the observed planar structures in which we are embedded are common and consistent with the standard $\Lambda$CDM model. 

\subsection{The probability of LGAs}

We find that the probability that a Local Group-like observer detects planar structures as prominent as ours 
is $0.3$--$0.4\%$. 
If we allow the local $9.3$Mpc disk and the larger $93$Mpc disk to be inclined by up to 15 degrees, then the probability increases
to about $1.5\%$. The probabilities are robust and vary little across three independent simulations with small variations in cosmological parameters.

One may ask whether the formation of these sheets is coupled to long-wavelength modes near the fundamental mode of the simulation box. 
As a baseline test we partitioned Uchuu and Flamingo-10K into eight sub-boxes of half the linear size, imposed periodic boundaries on each, 
and repeated the analysis (Appendix \ref{subbox}). The probabilities are slightly lower in the sub-boxes. The natural interpretation is that truncation at the artificial 
boundaries interrupts structures that extend across them. At the same time, the scatter among the eight realizations stays 5--20\% relative, and for the strictest $0^{\circ}$ configuration the standard deviation falls to 0.03--0.06 percentage points, so on scales of $1\,h^{-1}$Gpc the distribution of these structures is already close to homogeneous; the differences between Uchuu and the smaller MDPL2 volume in Table \ref{prob} are consistent with ordinary cosmic variance. 

\subsection{Properties of LGAs and large planar structures}

It is expected that LGAs are found nearer to clusters of galaxies than typical Pairs are, though the manner in which the geometry and abundance of clusters couple to the sheets remains unclear. 
Since the Mach number is expected to decrease with local overdensity \citep{Meriot_2022}, one might anticipate lower Mach numbers for LGAs. However, the proximity to clusters does not produce such a suppression.
We have repeated the Mach number comparison with lower halo mass thresholds 
of $3 \times 10^{11}\,M_{\odot}$ and $5 \times 10^{11}\,M_{\odot}$, as shown 
in Figure \ref{lowmass}; the agreement between LGAs and the general Pairs 
is unchanged, so this behavior is not sensitive to mass limits.

The most striking finding is the clustering of the planar regions. We find a 
correlation length of LGAs at roughly 60~Mpc, 
weakly dependent on inclinations between inner and outer disks and 
robust among simulations. 

There is a rough scaling argument against which to compare the correlation length. Empirically for clusters
of galaxies, the correlation length grows with richness in proportion to the mean separation,
$r_0 \approx 0.4\,d$ with $d = n^{-1/3}$ \citep{BahcallWest1992}. 
If one were to treat the planar structures represented by 
the LGAs as if they were in the same class of virialized structures, 
the abundances in Table~\ref{prob} would give $d \approx 200$~Mpc and $r_0 \approx 80$~Mpc
for the $15^\circ$ inclination, and $d \approx 320$--$370$~Mpc and $r_0 \approx 130$--$150$~Mpc for $0^\circ$.
When the local conditions are dropped and observers are required only to host the larger plane, the 
qualifying systems are about nine times as numerous as the LGAs and the scaling would predict
$r_0 \approx 40$~Mpc (Appendix~\ref{largeonly}). The measured correlation length nevertheless stays near 60~Mpc in
all three cases (Figures~\ref{consist} and \ref{xilarge}), despite the changes in system abundance by more than an order of magnitude. The clustering signal is thus dictated by the large planes and is impervious to further restriction imposed by requiring an aligned inner plane. The orientations of these systems provide interesting insight: a pair of large
planes has a tendency to be parallel at smaller separations, whereas
a pair of large planes tends toward random orientations at large separations (Figure~\ref{cospdf}).
Nearby markers thus often share the same planar structure, so their count overestimates the number of distinct structures. 

\bigskip

\begin{acknowledgments}
This work is supported by the National Key Research and Development Program of China and the Zhejiang provincial top level research support program. The analysis presented in this article
was carried out on the SilkRiver Supercomputer of Zhejiang University and 
the high-performance computing clusters `Starburst' and `Zimo' at the Purple Mountain Observatory, Chinese Academy of Sciences; we thank Lei Wang for his support in using them. We thank the Department of Astrophysical Sciences at Princeton University for hospitality during visits when much of this work was carried out.
We thank the Uchuu team and the Skies \& Universes site, the CosmoSim database at the Leibniz-Institute for Astrophysics Potsdam (AIP), and the Virgo Consortium for making the Uchuu, MultiDark, and FLAMINGO data available; we thank Joop Schaye and Matthieu Schaller for their assistance with the Flamingo-10K data. The MultiDark simulations were run on the GCS Supercomputer SuperMUC at LRZ with computing time provided by the Gauss Centre for Supercomputing and PRACE.
Finally, we are grateful to Andrey Kravtsov and Nick Gnedin for useful discussions.
\end{acknowledgments}

\software{numpy \citep{numpy2020},
          scipy \citep{scipy2020},
          matplotlib \citep{matplotlib2007},
          astropy \citep{astropy2022},
          h5py \citep{collette2013}}

\section*{Data Availability}

The simulation data underlying this work are accessible through their originating projects. 
The derived data products of this paper are available from the authors on reasonable request.
 
\appendix

\section{Simulation parameters and supplementary tests}\label{simpars}

Table \ref{para} lists the parameters of the datasets.

We first test the homogeneity and isotropy 
of the mass distribution. We select halos with mass over $10^{11}h^{-1}M_\odot$
and scatter their masses by 0.15~dex.
We compare the mass distribution along three box axes and against the uniform expectation. 
For a halo of mass $M_i$ whose coordinate along a given axis is $c_i$, the
mass-weighted cumulative distribution at coordinate $t$ is
\begin{equation}
    F(t)=\frac{\sum_i M_i\,\Theta(t-c_i)}{\sum_i M_i},
\end{equation}
where $\Theta$ is the step function. For the differential counterpart we divide
the box into slabs of thickness $\Delta t\simeq25$~Mpc spanning the full
cross-section, and define the fractional mass overdensity of the slab at $t$ as
\begin{equation}
    \delta(t)=\frac{M_{\rm slab}(t)}{\langle M\rangle}-1,
    \qquad \langle M\rangle=\frac{M_{\rm tot}}{N_{\rm slab}},
\end{equation}
which sums to zero over the box. We measure isotropy by comparing $\delta$ along the three axes and by the differences between axes, $\Delta F=F_A-F_B$ and $\delta_A-\delta_B$, and measure homogeneity by the departure of each axis from the uniform expectation, $\delta F=F-F_{\rm unif}$ with $F_{\rm unif}(t)=(t-t_{\min})/(t_{\max}-t_{\min})$.

\startlongtable
\begin{deluxetable*}{lccc}
     \tablewidth{0pt}
     \tablecaption{Simulation parameters\label{para}}
     \tablehead{
     \colhead{Parameter} & \colhead{Uchuu} & \colhead{MDPL2} & \colhead{Flamingo-10K}
     }
     \startdata
     box size & 2$h^{-1}$Gpc & 1$h^{-1}$Gpc & 2.8Gpc\\
     $N_{\rm part}$ & $12800^3$ & $3840^3$ & $10080^3$\\
     mass resolution & $3.27\times10^8h^{-1}M_{\odot}$ & $1.51\times10^9h^{-1}M_{\odot}$ & $5.72\times10^8h^{-1}M_{\odot}$\\
     initial redshift & 127 & 120 & / \\
     $h$ & 0.6774 & 0.6777 & 0.681 \\
     $\Omega_{\Lambda}$ & 0.6911 & 0.692885 & 0.694 \\
     $\Omega_{m}$ & 0.3089 & 0.307115 & 0.306 \\
     $\Omega_{b}$ & 0.0486 & 0.048206 & 0.0486\\
     $n_s$ & 0.9667 & 0.96 & 0.967 \\
     $\sigma_8$ & 0.8159 & 0.8228 & 0.807 \\
     subhalo type & Rockstar & Rockstar & HBT-HERONS \\
     \enddata
\end{deluxetable*}
\twocolumngrid

\begin{figure*}[!tp]
    \centering
    \includegraphics[width=\textwidth]{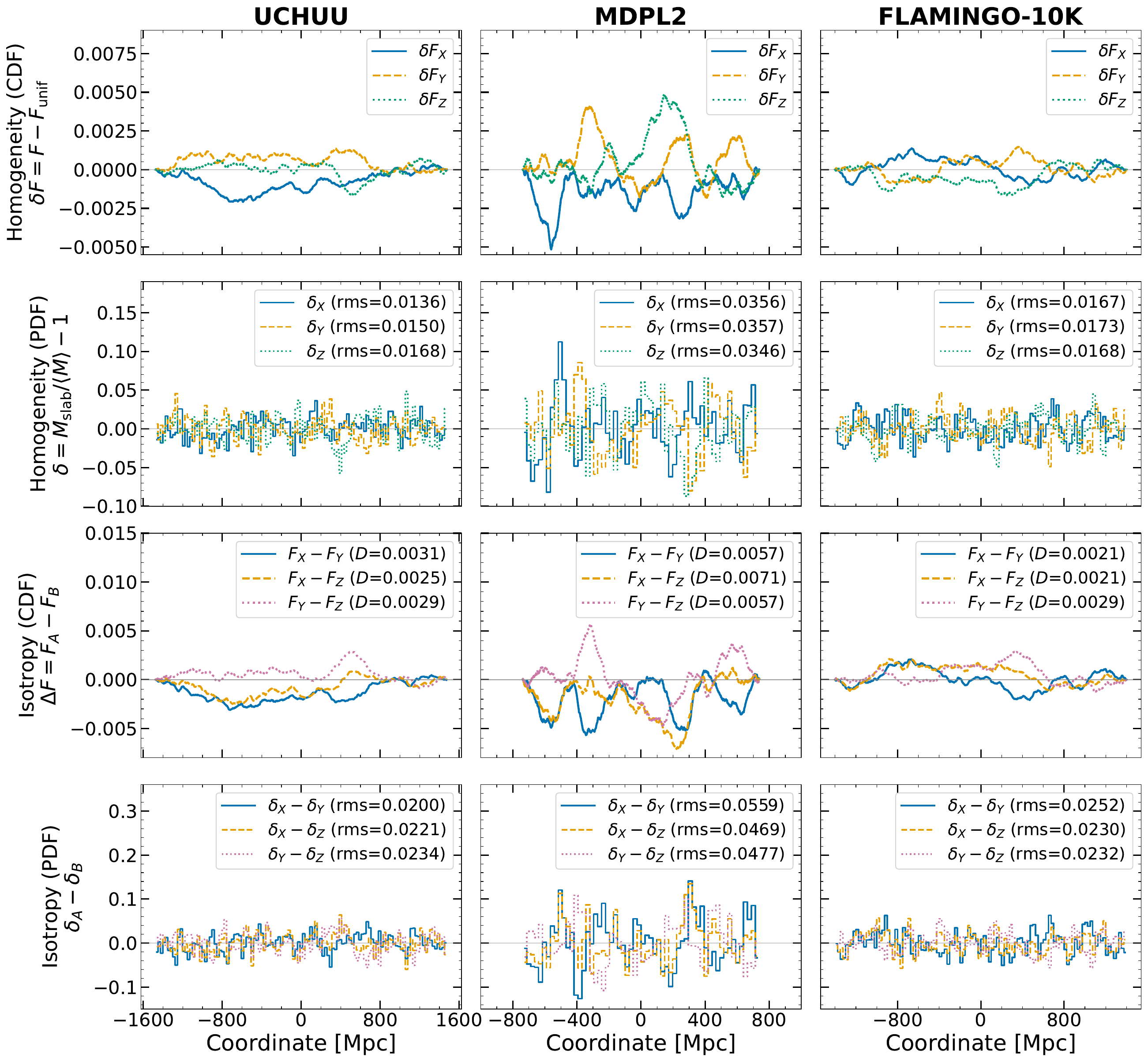}
\caption{Homogeneity and isotropy of the mass distribution in the three simulations (columns),
for halos above $10^{11}h^{-1}M_\odot$ with masses scattered by 0.15~dex. The horizontal axis is
the coordinate along a box axis.
Row 1 shows, for each axis $A$, $\delta F=F_A-F_{\rm unif}$, where
$F_A(t)=\sum_i M_i\,\Theta(t-c_{i,A})/\sum_i M_i$ is the fraction of halo mass with coordinate
$c_{i,A}<t$ along axis $A$, and $F_{\rm unif}(t)=(t-t_{\min})/(t_{\max}-t_{\min})$ is the
uniform expectation between the box ends $t_{\min}$ and $t_{\max}$.
Row 2 shows $\delta=M_{\rm slab}/\langle M\rangle-1$, the mass overdensity of slabs with 25~Mpc
thickness along each axis.
Rows 3 and 4 compare the axes in pairs: row 3 shows $\Delta F=F_A-F_B$, with the largest
difference $D=\max|\Delta F|$ given in the legend, and row 4 shows $\delta_A-\delta_B$.
The legends in rows 2 and 4 give the rms values over the slabs.}
    \label{isotropy}
\end{figure*}

Figure~\ref{isotropy} shows the results for the three simulations, plotted against the coordinate along a box axis.
Rows 2 and 4 test isotropy using the overdensity $\delta$ in 25~Mpc slabs along each axis and the difference $\delta_A-\delta_B$ between matched slabs along two axes. In each simulation, the three single-axis rms values $\sigma_X$, $\sigma_Y$, and $\sigma_Z$ (listed in the row 2 legends) are close to one another, ranging from 1.4\% to 1.7\% in Uchuu and Flamingo-10K and near 3.5\% in MDPL2. The MDPL2 value is roughly twice as large because its box is half as wide, giving a quarter of the slab cross-sectional area, and the slab variance scales inversely with that area. Under isotropy, the slab overdensities along the three orthogonal axes are independent realizations of a single zero-mean Gaussian distribution of width $\bar{\sigma}=[(\sigma_X^2+\sigma_Y^2+\sigma_Z^2)/3]^{1/2}$. Their covariance then vanishes, so each pairwise difference $\delta_A-\delta_B$ (row 4) follows a zero-mean Gaussian of width $\sqrt{2}\,\bar{\sigma}$. The measured values of $\sigma(\delta_A-\delta_B)$ in the row 4 legends lie between 0.93 and 1.12 times the expected $\sqrt{2}\,\bar{\sigma}$. We conduct one-sample Kolmogorov--Smirnov tests of each single-axis $\delta_A$ against $\mathcal{N}(0,\bar{\sigma}^2)$ and each difference $\delta_A-\delta_B$ against $\mathcal{N}(0,2\bar{\sigma}^2)$, which all yield $p>0.35$. The tail counts beyond $2\sigma$ also match the Gaussian expectation of 4.6\%. For Uchuu, MDPL2, and Flamingo-10K, the numbers of slabs with $|\delta_A|>2\bar{\sigma}$ are 14, 6, and 13, compared with expected counts of 16, 8, and 15, while the counts with $|\delta_A-\delta_B|>2\sqrt{2}\,\bar{\sigma}$ are 18, 10, and 14, again consistent. The cumulative differences $|\Delta F|$ in row 3 are likewise small, peaking at about 0.3\% of the total mass in Uchuu and Flamingo-10K and 0.7\% in MDPL2. None of the three boxes shows a preferred axis.

Row 1 tests spatial homogeneity through $\delta F$, the departure of the cumulative mass distribution from uniformity. Along each axis, $|\delta F|$ reaches at most about 0.2\% of the total mass in Uchuu and Flamingo-10K and 0.5\% in MDPL2, with no large-scale mass offset from one end of a box to the other. The most noticeable feature in row 1 is the dip in $\delta F_X$ for Uchuu about a quarter of the way across the box. To check whether this is an ordinary fluctuation, we divide each axis into four consecutive quarters, giving 12 segments per simulation, and compute the overall mass overdensity $\delta_q$ of each segment. Taking the rms of these 12 values as the fluctuation amplitude on this scale gives 0.30\%, 0.72\%, and 0.31\% for Uchuu, MDPL2, and Flamingo-10K. The Uchuu dip corresponds to the first quarter of the $X$ axis, where $\delta_q=-0.74\%$, or 2.5 times the rms. In 12 Gaussian draws, the probability that at least one exceeds this level is 14\% ($=1-0.9876^{12}$, where $0.9876$ is Gaussian fraction within 2.5$\sigma$). The largest departures in MDPL2 and Flamingo-10K are 2.4 and 1.7 rms, respectively. Across the three simulations, 1, 1, and 0 of the 12 segments lie beyond twice the rms, consistent with the Gaussian expectation of 0.55 per simulation. Therefore, the mass distributions in all three simulations are sufficiently homogeneous.

Similar to the considerations above, we examine further
whether the planar structures themselves exhibit any preferred orientation.
While the halo mass distribution itself is rather isotropic, the same may not necessarily be shared 
by large-scale structures. This is evaluated by analyzing
the statistics of the normal vectors of the planar structures associated with
the LGAs. For an isotropic directional distribution, the absolute projection
onto any arbitrary axis should satisfy $|\cos\theta|\sim\mathcal{U}(0,1)$, or
equivalently $\mathrm{PDF}(\theta)\propto\sin\theta$; we show both the
cumulative and the differential view of the projections onto three axes of the simulation box. Because the sample size of LGAs
under the $1\mathrm{mid}+1\mathrm{mid}+15^\circ$ criterion is relatively small, we use
the larger 1mid large-disk sample, which retains the 500/600-disk condition but
drops the local flat-pattern requirement and is shown to share the correlation length of the LGAs
(Appendix~\ref{largeonly}).

\begin{figure*}[!tp]
    \centering
    \includegraphics[width=\textwidth]{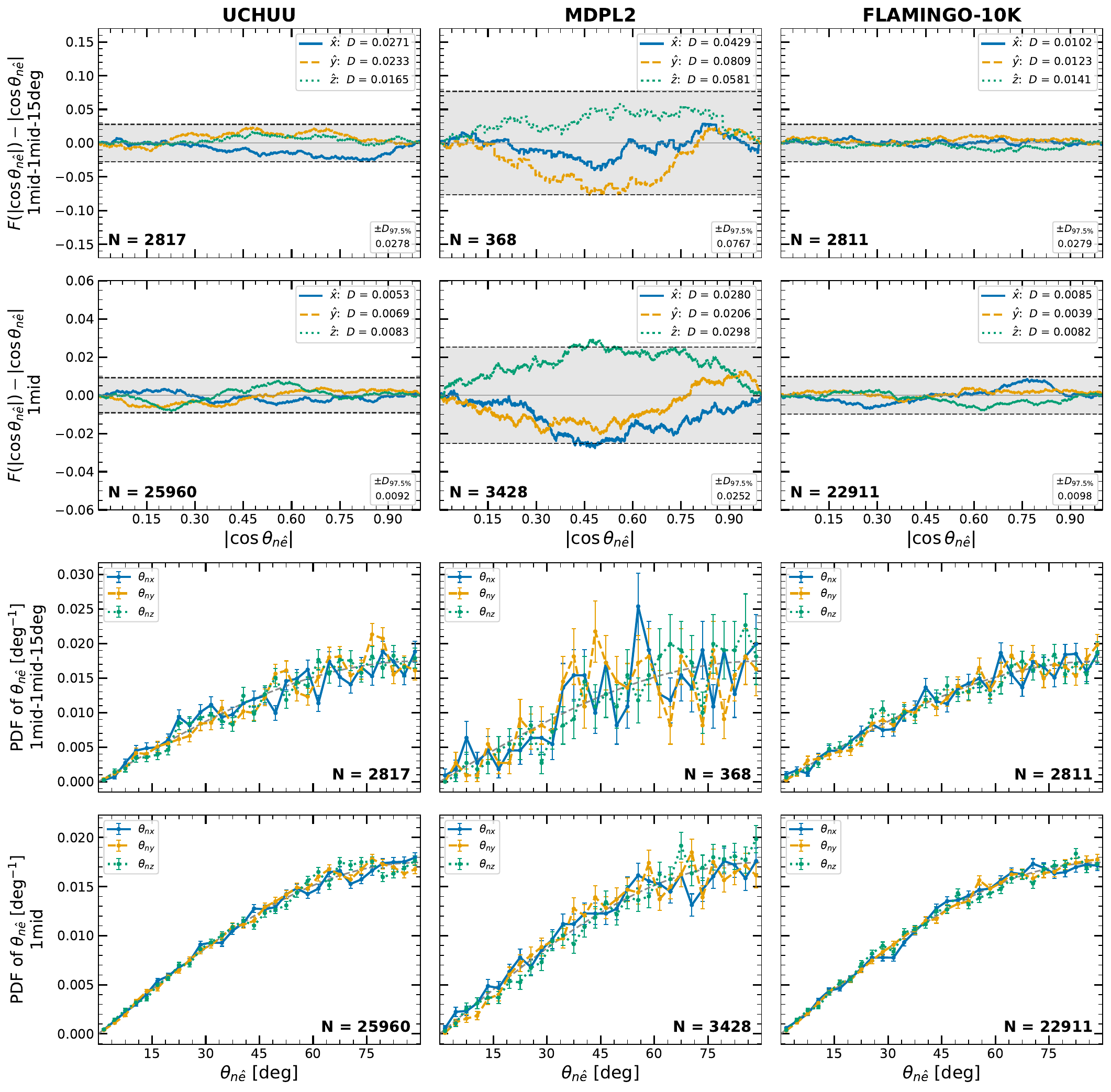}
\caption{Orientation preferences of planar structures. The first and third
rows show the distributions of the planar structure orientations projected
onto the Cartesian axes for LGAs under the $1\mathrm{mid}+1\mathrm{mid}+15^\circ$ configuration; the
second and fourth rows show the distributions for halos requiring only the
1mid Large Flat Structure condition. Rows 1 and 2 display the deviations of the CDF
of the projected cosines $c_{e}=|\mathbf{n}\cdot\hat{e}|$ from the isotropic
expectation, whereas rows 3 and 4 present the distributions of the
inclination angle $\theta$ between the planar-structure normal and the
coordinate axes. In rows 1 and 2,
$D$ represents the maximum absolute difference between the empirical
CDF and the isotropic expectation, and $D_{97.5\%}$ is the
97.5th percentile of the one-sample KS statistic at
that sample size, with the gray shaded band bounded by dashed lines marking
the corresponding acceptance region. In rows 3 and 4, the dashed lines
indicate the reference curve for an isotropic angular distribution
($\propto\sin\theta$), with Poisson error bars on the curves. The lower
corners indicate the total sample sizes used.}
    \label{normal_theta}
\end{figure*}

As shown in Figure~\ref{normal_theta}, most projections 
remain safely bounded inside the $\pm D_{97.5\%}$ band in both samples except MDPL2;
MDPL2 is the smallest box and has the largest slab fluctuations, so we read 
these excursions as finite-box fluctuations.
In all six PDF panels, the projections track the isotropic $\sin\theta$ reference
within the Poisson errors over the full range. Even in the small sample scenario the 
consistency is good.
Therefore, no evidence points to
an orientation preference for planar structures.

\section{Clustering of Distant-Disk samples}\label{largeonly}

The comparison in Section \ref{discussions} between the abundance of the LGAs and their correlation length interprets each LGA as a marker of one distinct distant-disk. 
To test this interpretation, we implement two checks.

\begin{figure}[!htbp]
    \centering
    \includegraphics[width=\linewidth]{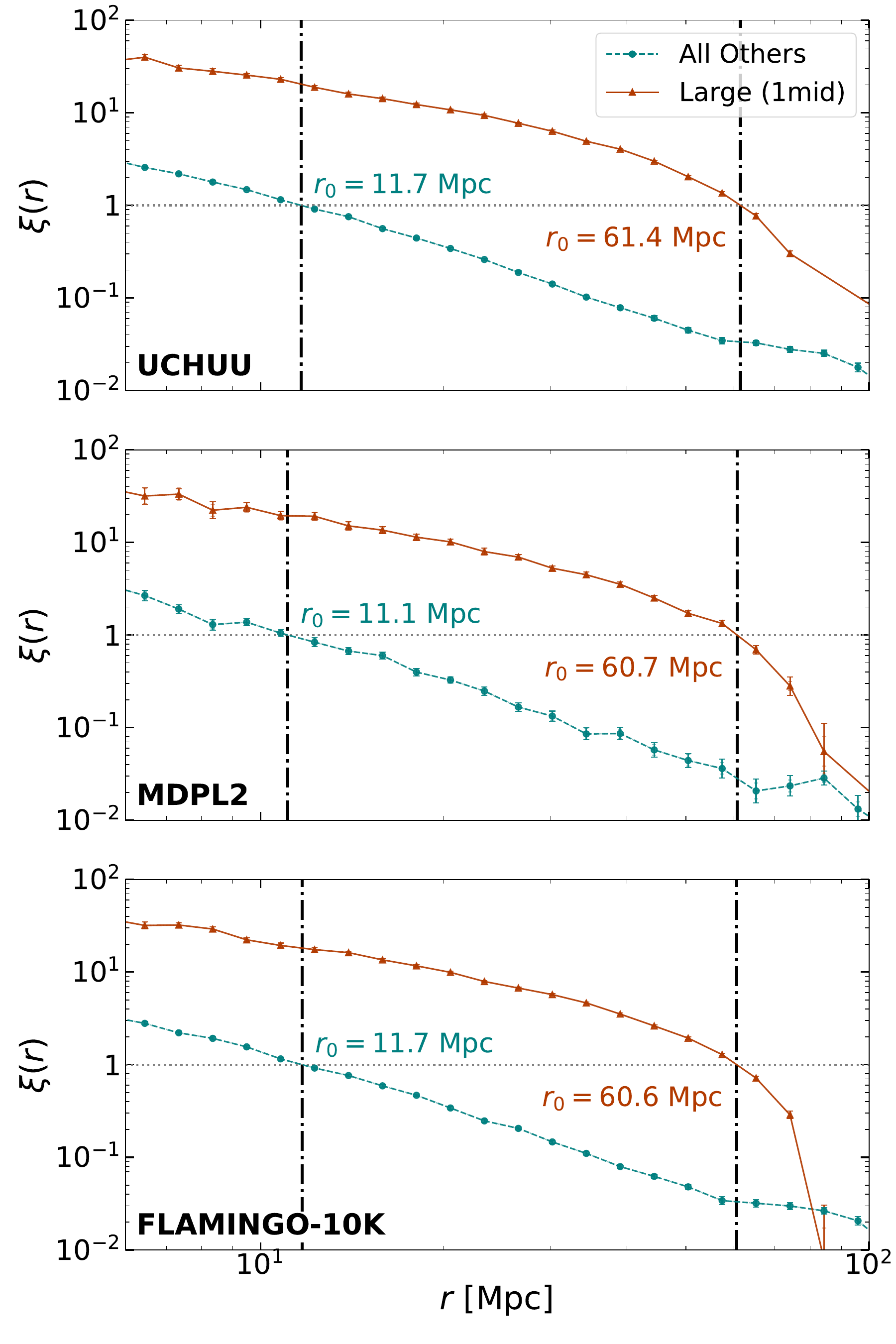}
    \caption{Two-point correlation functions of the 1mid large-disk sample (solid; counts in Column (7) of Table~\ref{prob}) and of Pairs in the valid sample without a Large Flat Structure (dashed), for the three simulations. Error bars are from 27-fold jackknife resampling, as in Figure \ref{corr}.}
    \label{xilarge}
\end{figure}

First we relax the selection. We keep conditions 1 and 2 of Section \ref{LGA} to get a valid observer 
hosting a Large Flat Structure, and drop the local-disk condition and the angular coupling. The counts 
of this 1mid `large-disk' sample are those of Column (7) in Table \ref{prob}, about nine times those of 
the LGAs; the mean separation is $d \approx 100$~Mpc, for which the scaling of Section \ref{discussions} 
would predict $r_0 \approx 40$~Mpc.
Figure \ref{xilarge} shows the two-point correlation function of this sample.
The measured correlation length is about 60~Mpc, indistinguishable 
from that of the LGAs and far from the scaling prediction.

\begin{figure}[!htbp]
    \centering
    \includegraphics[width=\linewidth]{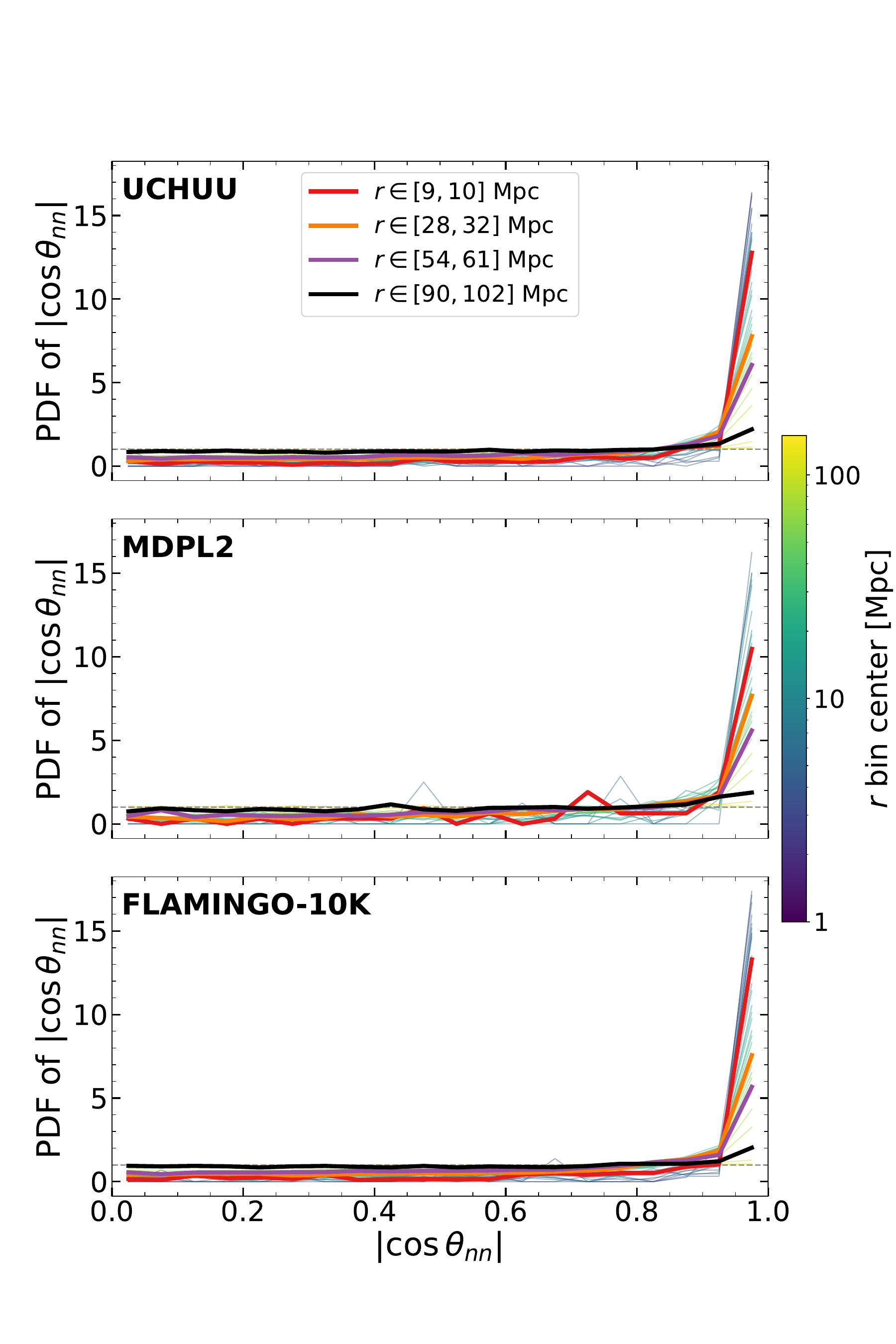}
    \caption{Distributions of the relative orientations of systems in the 1mid
    large-disk sample. For any two systems in a 
    distance bin of the correlation function that we compute, 
    $|\cos\theta_{nn}|$ is the absolute cosine of the angle between the two plane normals. 
    Thin curves span all distance bins, colored by bin center, and four representative bins 
    are highlighted. The horizontal dashed line at unity marks the expectation for 
    random orientations.}
    \label{cospdf}
\end{figure}

Second we measure the relative orientations between systems in the large-disk sample. We take for each system 
the normal of the 600-disk as its direction and for any two systems whose separation falls in a given distance bin of the correlation function we record the dot product of the two direction vectors, 
$|\cos\theta_{nn}|$. 
Figure \ref{cospdf} shows the normalized probability densities of $|\cos\theta_{nn}|$ bin by bin. 
At separations below $\sim$10~Mpc the normals are strongly aligned. The alignment weakens 
with separation, and by $r \approx 100$~Mpc the distributions are nearly flat. 

\begin{figure*}[!tp]
    \centering
    \includegraphics[width=\textwidth]{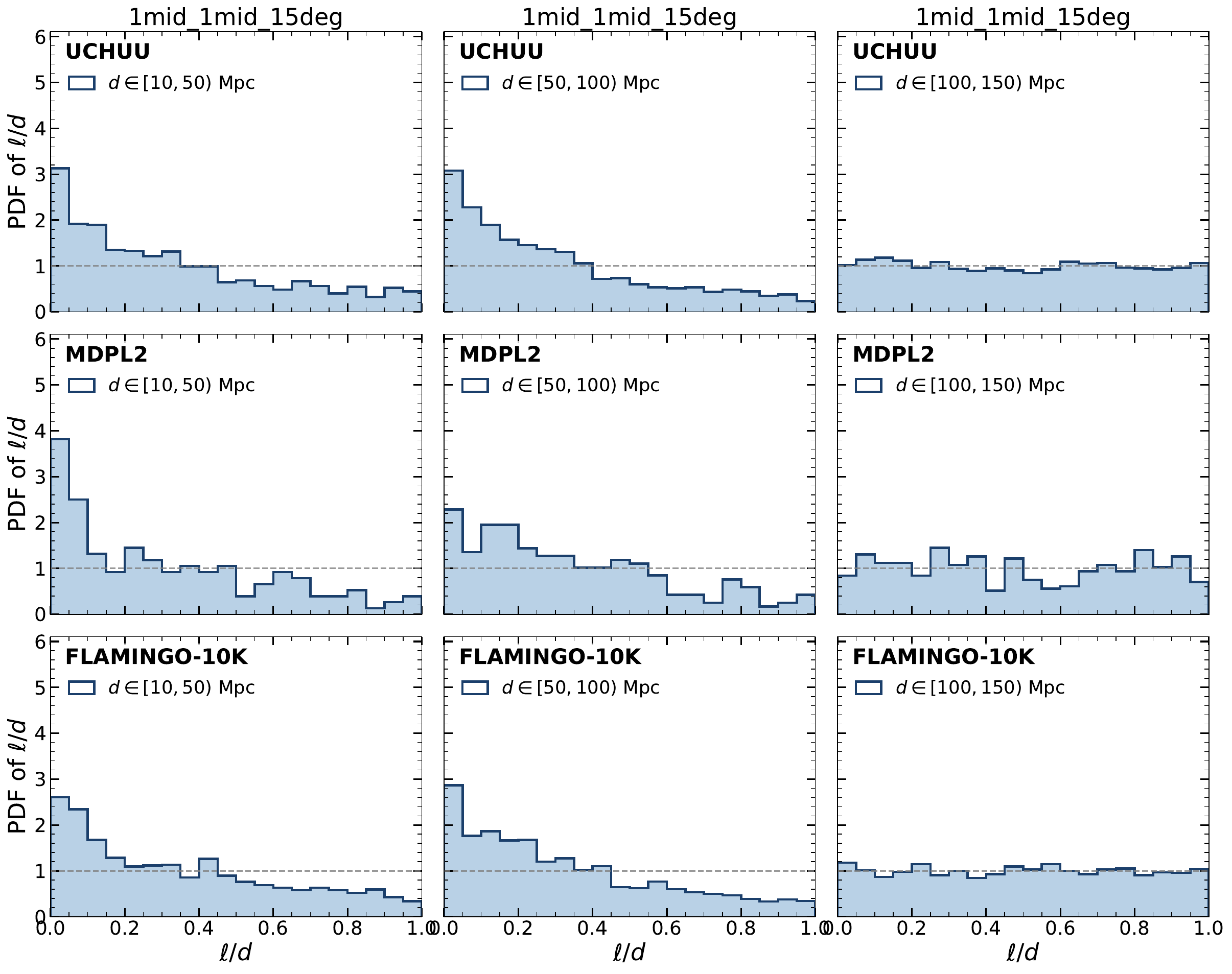}
    \caption{Probability distributions of $l/d$ for the $1\mathrm{mid}+1\mathrm{mid}+15^\circ$ configuration. The left, middle, and right panels show the PDFs for two LGAs separated by $10\text{--}50\text{ Mpc}$, $50\text{--}100\text{ Mpc}$, and $100\text{--}150\text{ Mpc}$, respectively; the gray dashed line indicates the uniform distribution over $(0, 1)$.}
    \label{ld}
\end{figure*}

To indirectly assess the actual abundance of planar structures, a useful test is to examine the distribution of $l/d$, where $d$ is the separation between any two LGAs and $l$ is the normal distance from one LGA to the disk of the other. A value of $l/d$ approaching 0 signifies that the two LGAs reside within the same planar structure, whereas $l/d \to 1$ indicates that one LGA lies entirely outside the plane of the other, tending toward a parallel-stacked configuration. Under the assumption of isotropy, $l/d$ follows a uniform distribution over $(0, 1)$. Figure \ref{ld} presents the $l/d$ distributions across different separation bins. When the separation is in the range of $10\text{--}100\text{ Mpc}$ (left and middle panels), an excess in low $l/d$ values appears, indicating that LGAs within $100\text{ Mpc}$ of each other tend to share the same plane. This together with Figure \ref{cospdf} further indicates that LGAs oversample planar structures. Once the separation exceeds $100\text{ Mpc}$ (right panel), the uniformity is recovered, and the plane-based correlation among LGAs disappears.

\section{Halo mass conversion for Flamingo-10K} \label{conver}

Flamingo-10K uses a different halo finder, so we treat it as a reference for testing the sensitivity of our results rather than as part of a strictly homogeneous set. For this purpose we compute $M_{102c}$, the virial mass at $z=0$ derived by \citet{bryanStatisticalPropertiesXRay1998},
\begin{equation}
    \Delta_{vir}=18\pi^2+82(\Omega(z)-1)-39[\Omega(z)-1]^2\,.
\end{equation}
With the FLAMINGO D3A cosmological parameters (specifically $\Omega_{m,0}=0.306$) the virial overdensity is $\Delta_{vir}\approx102$. We employ the empirical concentration-mass-redshift relation of \citet{ludlowMassConcentrationRedshift2016},
\begin{equation}
    c(\nu)=c_0(\frac{\nu}{\nu_0})^{-\gamma_1}[1+(\frac{\nu}{\nu_0})^{1/\beta}]^{-\beta(\gamma_2-\gamma_1)}\,,
\end{equation}
and model each halo with the Navarro-Frenk-White (NFW) density profile \citep{1996ApJ...462..563N,Navarro_1997}, from which
\begin{equation}
    \begin{aligned}
        M_{200c} &= \frac{4\pi}{3}\,200\,\rho_{\rm crit}\,(r_{200})^3 = 4\pi\,\rho_s\,r_s^3\, f(c)\\
        M_{102c} &= \frac{4\pi}{3}\,102\,\rho_{\rm crit}\,(r_{102})^3 = 4\pi\,\rho_s\,r_s^3\, f(c_{102})\,.
    \end{aligned}
\end{equation}
For a given halo the scale radius $r_s$ and scale density $\rho_s$ are constant, so the mass ratio between overdensity definitions is
\begin{equation}
    \frac{M_{102c}}{M_{200c}} = \frac{f(c_{102})}{f(c)}\,,
\end{equation}
where the auxiliary function
\begin{equation}
    f(x)=\ln(1+x)-\frac{x}{1+x}
\end{equation}
characterizes the NFW mass profile. The concentration parameter $c_{\Delta} = r_{\Delta}/r_s$ relates the virial radius to the scale radius, and combining these relations yields an implicit equation for the concentration ratio,
\begin{equation}
    \frac{c_{102}}{c} = \frac{r_{102}}{r_{200}} = \left(\frac{200}{102}\cdot \frac{f(c_{102})}{f(c)}\right)^{1/3}\,,
\end{equation}
which we solve numerically by fixed-point iteration of the right-hand side 
with an absolute tolerance of 0.1. 
The resulting $M_{102c}$ exceeds $M_{200c}$ by approximately 10--30\%
consistently across the halo sample.

\section{Asymptotic probabilities}\label{asym}

\begin{figure*}[!tp]
    \centering
    \includegraphics[width=\textwidth]{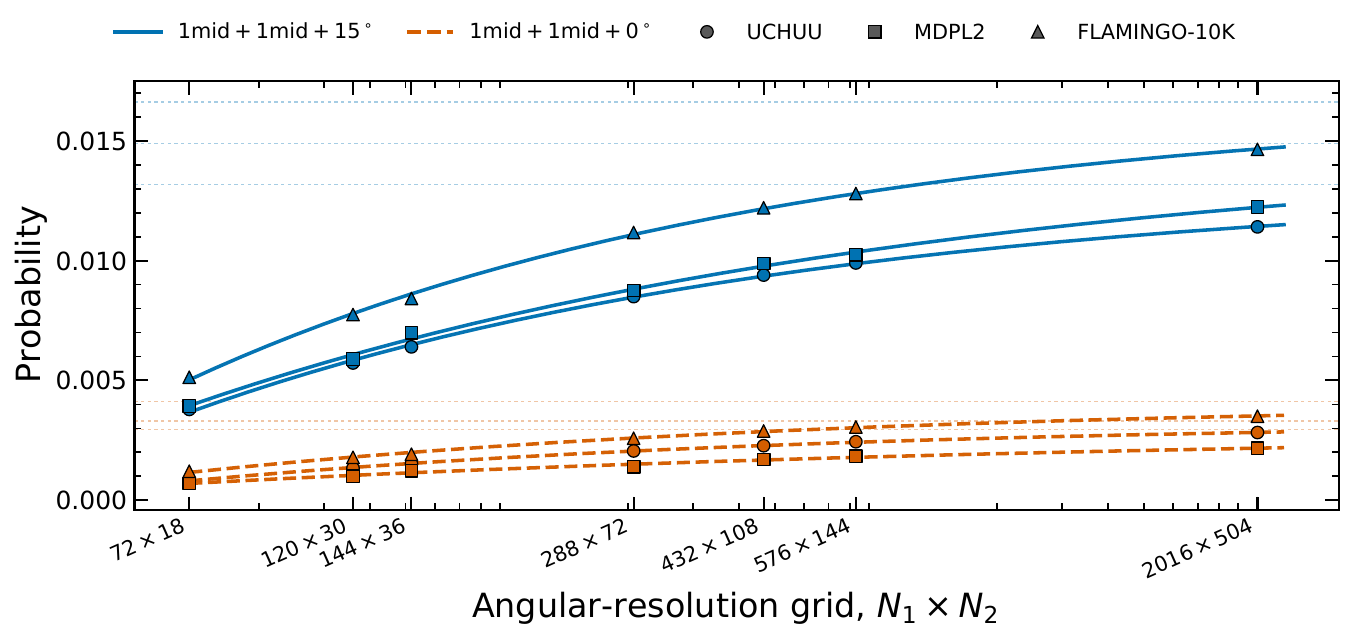}
    \caption{LGA probabilities at a sequence of discrete sampling resolutions. Curves are power-law fits obtained with logarithmic weighting; horizontal dotted lines mark the constant asymptotes derived from the fits.}
    \label{asymfigure}
\end{figure*}

The probability of identifying planar structures is defined in angular space, and its estimation is necessarily limited by the density of angular sampling. 
The comparative tests in the main text rely on a resolution of $\approx 10^6$ samples per observer, which we find sufficient for a consistent baseline. 
A finer angular grid is expected to be more sensitive to subtler alignments, so it is helpful to consider the limiting distribution as the sampling approaches continuum. To this end we have computed the frequencies across resolutions from $\sim 10^3$ to $\sim 10^6$ samples. The resulting probabilities are well described by a power law with a constant asymptote,
\begin{equation}
P(N) = P_{\infty} + A \cdot N^{-\gamma}\,,
\label{asympt:eq}
\end{equation}
where $P_{\infty}$ is the limiting probability as $N \rightarrow \infty$. To mitigate noise from small samplings we adopt a logarithmic weighting for the fit,
\begin{equation}w(N) = 1 + 4\times \left( \frac{\log_{10}N - \log_{10}N_{min}}{\log_{10}N_{max} - \log_{10}N_{min}} \right)^{1.5}\,,
\end{equation}
where $N$ is the number of samples. For the fiducial $1\mathrm{mid}+1\mathrm{mid}+15^\circ$ 
configuration the occurrence probability is enveloped by $1.5\%$ (not including 
Flamingo-10K). The fits are shown in Figure \ref{asymfigure} 
and Table \ref{asymtable}. 
The slightly higher values in Flamingo-10K
reflect the lower effective mass threshold associated with 
the $M_{102c}$ definition.

We make no guarantee that this trend has reached numerical convergence. However, with $\sim 10^6$ angular samplings the recovered value is closely comparable to the observational result. The same sampling density ensures that the 500-disk solid-angle fractions entering the LGA criteria are not systematically underestimated by the discreteness of the grid. 
At $N\approx 10^6$, the measured probabilities are slightly below the asymptotes $P_\infty$ (Figure~\ref{asymfigure}); therefore, in the main text we report the extrapolated $P_\infty$ values as conservative upper bounds.

\begin{deluxetable*}{llrrrrr}
\tabletypesize{\scriptsize}
\tablewidth{0pt}
\tablecaption{Parameters of the power-law fits for the two configurations\label{asymtable}}
\tablehead{
\colhead{Simulation} &
\colhead{Configuration} &
\colhead{$P_{\infty}$} &
\colhead{$A$} &
\colhead{$\gamma$} &
\colhead{max $|\Delta P|$} &
\colhead{RMS $\Delta P$} \\
\hline
\colhead{(1)} & \colhead{(2)} & \colhead{(3)} & \colhead{(4)} &
\colhead{(5)} & \colhead{(6)} & \colhead{(7)}
}
\startdata
  Uchuu & $1\mathrm{mid}+1\mathrm{mid}+15^\circ$ & 0.013179 & $-0.058768$ & 0.2541 & $1.19\times10^{-4}$ & $7.53\times10^{-5}$ \\
        & $1\mathrm{mid}+1\mathrm{mid}+0^\circ$  & 0.003307 & $-0.014858$ & 0.2485 & $2.84\times10^{-5}$ & $1.81\times10^{-5}$ \\
\hline
  MDPL2 & $1\mathrm{mid}+1\mathrm{mid}+15^\circ$ & 0.014897 & $-0.050020$ & 0.2119 & $2.55\times10^{-4}$ & $1.36\times10^{-4}$ \\
        & $1\mathrm{mid}+1\mathrm{mid}+0^\circ$  & 0.002943 & $-0.007062$ & 0.1599 & $1.04\times10^{-4}$ & $5.61\times10^{-5}$ \\
\hline
  Flamingo-10K & $1\mathrm{mid}+1\mathrm{mid}+15^\circ$ & 0.016613 & $-0.078853$ & 0.2675 & $1.92\times10^{-4}$ & $9.30\times10^{-5}$ \\
               & $1\mathrm{mid}+1\mathrm{mid}+0^\circ$  & 0.004119 & $-0.016487$ & 0.2395 & $7.61\times10^{-5}$ & $3.83\times10^{-5}$ \\
\enddata
\tablecomments{Column (3) lists the constant asymptotes for the LGA probabilities; columns (4) and (5) give the estimated parameters of Equation \ref{asympt:eq}. Column (6) is the maximum deviation of the data points from the fitted curve; column (7) is a measure of the typical scatter.}
\end{deluxetable*}

\section{Low mass cosmic Mach number profiles}\label{app-lowmass}

\begin{figure*}[!tp]
    \plotone{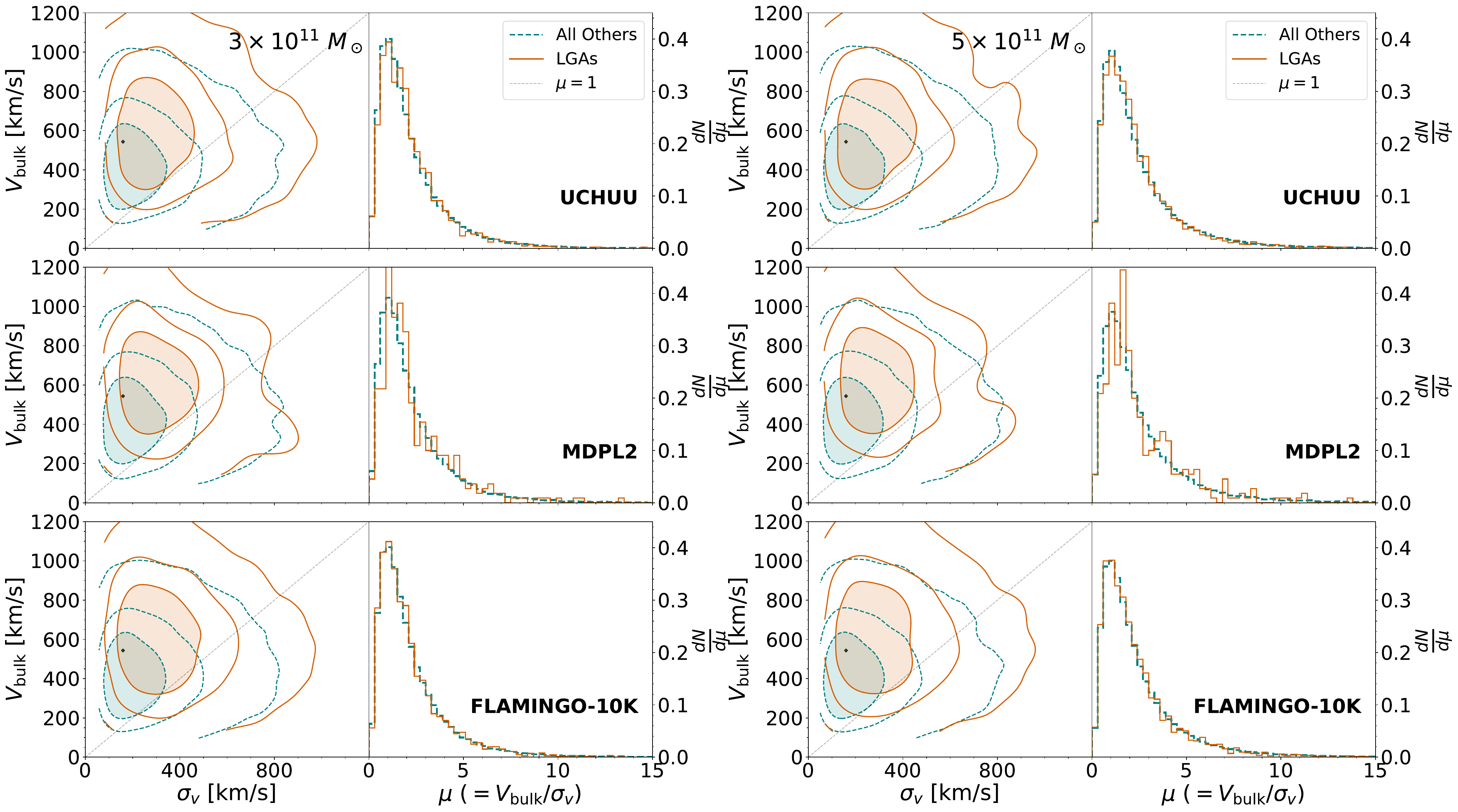}
    \caption{Bulk flow velocity against velocity dispersion and cosmic Mach numbers including less massive halos in local environments. 
    The left column includes the analysis for halos more massive than $3\times10^{11}M_{\odot}$, 
    and the right column includes halos more massive than $5\times10^{11}M_{\odot}$.
    Other settings are the same as Figure \ref{fig:env}(c) and (d).}
    \label{lowmass}
\end{figure*}

We include sensitivity tests for the cosmic Mach number with different lower halo mass thresholds. Comparing Figure \ref{lowmass} with the (c), (d) panels of Figure \ref{fig:env}, either lower threshold gives approximately the same distributions of both bulk velocity and $\sigma_{v}$, which indicates that in terms of local dynamics, lower-mass galaxies behave similarly to more massive ones.

\section{Sub-box statistics}\label{subbox}

To construct the sub-box realizations discussed in Section \ref{discussions}, we shift the comoving coordinates of all particles within the parent volume so that the system is centered at $(0,0,0)$, and assign particles to sub-boxes by the octants they occupy; the notation `$--+$' designates the sub-volume with $(x<0,\,y<0,\,z\ge0)$. Periodic boundary conditions are then imposed on each sub-box, and the full statistical analysis is repeated. Table \ref{prob_summary} summarizes the eight realizations for each configuration, with columns as in Table \ref{prob}. Beyond the modest suppression discussed in the main text, the scatter among the eight realizations amounts to a relative error of 5--10\% for the $1\mathrm{mid}+1\mathrm{mid}+15^\circ$ configuration and 10--20\% for $1\mathrm{mid}+1\mathrm{mid}+0^\circ$.

\startlongtable
\begin{deluxetable*}{llrrrrrrr}
\tabletypesize{\scriptsize}
\tablewidth{0pt}
\tabcolsep=3pt
\tablecaption{Summary of LGA Measurements over Eight Sub-box Realizations \label{prob_summary}}
\tablehead{
\colhead{Configuration} &
\colhead{Simulation} &
\colhead{Sample} &
\colhead{Valid} &
\colhead{Local} &
\colhead{600-Disk} &
\colhead{Large} &
\colhead{LGA} &
\colhead{Fraction} \\
\colhead{} &
\colhead{} &
\colhead{Size} &
\colhead{Sample} &
\colhead{Disk} &
\colhead{Existence} &
\colhead{Disk} &
\colhead{} &
\colhead{(\%)} \\
\hline
\colhead{(1)} & \colhead{(2)} & \colhead{(3)} & \colhead{(4)} &
\colhead{(5)} & \colhead{(6)} & \colhead{(7)} & \colhead{(8)} & \colhead{(9)}
}
\startdata
$1\mathrm{mid}+1\mathrm{mid}+15^\circ$ & Uchuu        & $39682 \pm 266$ & $30726 \pm 214$ & $16294 \pm 158$ & $3539 \pm 184$ & $3094 \pm 184$ & $332 \pm 34$ & $1.082 \pm 0.110$ \\
                                       & Flamingo-10K & $30910 \pm 254$ & $24015 \pm 277$ & $13007 \pm 159$ & $3062 \pm 150$ & $2663 \pm 149$ & $319 \pm 18$ & $1.326 \pm 0.066$ \\
\hline
$1\mathrm{mid}+1\mathrm{mid}+0^\circ$  & Uchuu        & $39682 \pm 266$ & $30726 \pm 214$ & $16294 \pm 158$ & $3539 \pm 184$ & $3094 \pm 184$ & $86 \pm 18$ & $0.278 \pm 0.058$ \\
                                       & Flamingo-10K & $30910 \pm 254$ & $24015 \pm 277$ & $13007 \pm 159$ & $3062 \pm 150$ & $2663 \pm 149$ & $79 \pm 8$ & $0.329 \pm 0.033$ \\
\enddata
\tablecomments{For each configuration, the quoted values are the mean and standard deviation over 
the eight sub-box realizations (\texttt{---}, \texttt{--+}, \texttt{-+-}, \texttt{-++}, \texttt{+--}, 
\texttt{+-+}, \texttt{++-}, \texttt{+++}).}
\end{deluxetable*}
\twocolumngrid

\clearpage
\bibliography{ref}
\bibliographystyle{aasjournalv7}

\end{document}